\documentclass[aps,pra,reprint,showpacs,floatfix,superscriptaddress]{revtex4-1}
\usepackage{graphicx}
\usepackage{amsmath,amsfonts,amscd, physics}
\usepackage{inputenc}
\usepackage{mathrsfs, dsfont}
\usepackage{mathtools}
\usepackage{stmaryrd}
\usepackage{array}
\usepackage{bm}
\usepackage{setspace}
\usepackage[T1]{fontenc}
\usepackage{txfonts}
\usepackage{lineno}
\usepackage{csquotes}
\usepackage{xcolor}

\begin{document}

\title{Quantum advantage of time-reversed ancilla-based metrology of absorption parameters}
\author{Jiaxuan Wang}
\affiliation{Department of Physics and Astronomy, Texas A\&M University, College Station, Texas 77843, USA}
\affiliation{Institute for Quantum Science and Engineering, Texas A\&M University, College Station, Texas 77843, USA}
\author{Ruynet. L. de Matos Filho}
\affiliation{Instituto de F\'isica, Universidade Federal do Rio de Janeiro, Rio de Janeiro, RJ 21941-972, Brazil}
\author{Girish S. Agarwal}
\affiliation{Department of Physics and Astronomy, Texas A\&M University, College Station, Texas 77843, USA}
\affiliation{Institute for Quantum Science and Engineering, Texas A\&M University, College Station, Texas 77843, USA}
\affiliation{Department of Biological and Agricultural Engineering, Texas A\&M University, College Station, Texas 77843, USA}
\author{Luiz Davidovich}
\affiliation{Department of Physics and Astronomy, Texas A\&M University, College Station, Texas 77843, USA}
\affiliation{Institute for Quantum Science and Engineering, Texas A\&M University, College Station, Texas 77843, USA}
\affiliation{Instituto de F\'isica, Universidade Federal do Rio de Janeiro, Rio de Janeiro, RJ 21941-972, Brazil}

\begin{abstract} 
Quantum estimation of parameters defining open-system dynamics may be enhanced by using ancillas that are entangled with the probe but are not submitted to the dynamics.  Here we {consider} the important problem of estimation of transmission of light by a sample, with losses due to absorption and scattering. We show, through the determination of the quantum Fisher information, that the ancilla strategy leads {to the best possible precision in single-mode estimation} -- the one obtained for a  Fock state input --, through joint photon-counting of probe and ancilla, which are modes of a bimodal squeezed state {produced by an optical parametric amplifier}.  This proposal overcomes the challenge of producing and detecting high photon-number Fock states, {and it is quite robust against additional noise: we show that it is immune to phase noise and the precision does not change {if the incoming state gets disentangled.} } {Furthermore, the quantum gain is still present under moderate photon losses of the input beams.} We also discuss an alternative to joint photon counting, which is readily implementable with present technology, and approaches the quantum Fisher information result for weak absorption, {even with moderate photons losses of the input beams before the sample is probed}: a time-reversal procedure, placing the sample between two optical parametric amplifiers, with the second undoing the squeezing produced by the first one. The precision of estimation of the loss parameter is obtained from the average outgoing total photon number and its variance.  In both procedures, the state of the probe and the detection procedure are independent of the value of the parameter. 

\end{abstract}

\maketitle

\section{Introduction}
Quantum sensing involves the use of quantum resources, like entanglement and squeezing, for the estimation of parameters characteristic of a physical process, through its action on a probe that, upon a proper measurement, allows the estimation of the value of the parameters \cite{helstrom,holevo},  It has become one of the most active areas of quantum information, with important theoretical developments and useful devices \cite{paola}. Entanglement of the probe with an ancilla that is not submitted to the physical process may increase the precision of estimation \cite{fugiwara1,fugiwara2}. This is true, however, only to open-system dynamics. Here we {apply the ancilla protocol to the estimation of the photon-loss coefficient of a sample, due to absorption and scattering of light.} The probe and the ancilla correspond to two modes of a bimodal squeezed state, produced by an optical parametric amplifier (OPA). {Relevant aspects of the ancilla protocol for estimating absorption were studied theoretically {\cite{monras,nair1,gong}} and experimentally \cite{moreau,losero}. Here we show, through a novel and clarifying analytical procedure, that, for a given input intensity through the sample, this scheme leads to precision of estimation identical {to the best possible one for single-mode estimation}, obtained by using  Fock states {\cite{paris2,adesso,gammaT}.} This has the advantage of avoiding the preparation of Fock states with high photon numbers, {though in principle such states can be heralded from a two-mode squeezed vacuum state via the use of photon-number resolving detectors, as demonstrated in \cite{sturges} for up to five photons}.  {Our derivation allows us to determine the corresponding best measurement: a joint photon-counting of the outgoing probe and the ancilla.} Since this measurement could be challenging with current technologies, except for joint Fock-state spaces of very small size, we present a time-reversal detection alternative, consisting in placing the sample between two bimodal squeezing transformations {(two OPAs)}, such that the second squeezing is the inverse of the first one.   The precision in the estimation of the loss parameter is obtained from the averaged total photon number and its variance after the second transformation. 

Time-reversal has been shown to increase the precision of estimation of parameters characterizing unitary processes, beyond the classical limit, like displacements  \cite{toscano,penasa,burd,agarwal} and phases {\cite{agarwal,macri,nolan,vuletic,linnemann}}. For absorption estimation,  time-reversal must be complemented by the use of ancilla. We show that, for small absorption, the estimation obtained with this approach is very close to the best possible precision, obtained from single-mode probes prepared in a Fock state \cite{adesso,gammaT}, and it is superior to proposals based on a single probe (no ancilla), prepared in a squeezed state \cite{paris2}.  It has the further advantage that neither the input state nor the detection procedure depends on the value of the parameter, which simplifies the experimental realization, {avoiding resource-consuming adaptive measurements}, and should motivate useful applications.  {While adaptive strategies require additional measurements, they can be useful in several situations \cite{armen,wheatley,berni,lovett}, and specially when the number of probes is small \cite{hentschel}. Our method, has the advantage of avoiding the adaptation of the apparatus throughout the measurement. } {We show that quantum advantage is still present under moderate photon losses of the input beam.}

Measuring the probe, after it undergoes the parameter-dependent dynamics, leads to an estimation of the parameter through a function -- an estimator -- that maps an experimental data set to a possible value of the parameter. There are four basic questions that one would like to answer: (i) How to define the precision of the estimate?; (ii) How to get the precision from the experimental results?; (iii) What is the best initial state of the probe, in order to get the best precision?; and (iv) What is the best measurement procedure?

For unbiased estimations, the average of the estimator over a large number of realizations of the measurement coincides with the true value of the parameter. In this case, the precision of the estimation may be quantified by the standard deviation of the measured values of the parameter with respect to the average: $\Delta X=\sqrt{\langle X^2\rangle-\langle X\rangle^2}$. Within the classical framework, a lower bound for the variance was obtained by Cram\'er and Rao \cite{cramer}, and shown by Fisher \cite{fisher} to be attainable when the distribution of the possible values of the parameter is Gaussian, or when the number of repetitions of the measurement is much larger than one. The Cram\'er-Rao bound is expressed in terms of the {\it Fisher information},
\begin{equation}\label{fisher}
F(X)=\sum_j\frac{1}{P_j(X)}\left[\frac{dP_j(X)}{dX}\right]^2\,,
\end{equation}
where $P_j(X)$ is the probability of getting an experimental result $j$ if the value of the parameter is $X$. One has then $\Delta X\ge1/\sqrt{{\cal N}F(X)}$, where ${\cal N}$ is the number of independent measurements.

Generalization of this early work to quantum mechanics, through maximization of $F(X)$ over all possible quantum measurements, leads to the inequality
\begin{equation}\label{qfi}
\Delta X\ge1/\sqrt{{\cal N}{\cal F}_Q(X)}\,,
\end{equation}
where ${\cal F}_Q(X)$ is the quantum Fisher information (QFI).   This relation implies that the precision in the estimation of parameters can be increased beyond the minimum uncertainty obtained by classical means, usually referred to as the standard limit \cite{helstrom,holevo,braunstein2}. Quantum advantage has been proven for estimations of displacements or rotations \cite{helstrom0,caves,walborn,mason,burd,gilmore,agarwal}, phases \cite{holevo0,monras1,paris,giovannetti,rafa,bruno,roccia}, electromagnetic fields \cite{mason,gilmore,latune,penasa,facon}, damping and temperature \cite{paris2,adesso,gammaT},  the gravitational field \cite{gillies,grav,westphal}, or yet the  squeezing parameter of electromagnetic fields \cite{milburn,chiribella}. More recently, interesting applications have been demonstrated, among them gravimeters \cite{stray}, accelerometers \cite{yu},  gyroscopes \cite{gyro},  magnetometers \cite{magnetometer}, high-resolution spectroscopy \cite{hr}, detection of gravitational waves \cite{caves,ligo}, and ultra-precise atomic clocks \cite{clock}. Quantum metrology also concerns conceptual questions related to foundations of quantum mechanics, as, for instance, the meaning of number-phase and energy-time uncertainty relations \cite{braunstein}, this last one being related to the quantum speed limit \cite{anandan,taddei}. 

For noiseless quantum processes, with probe dynamics governed by unitary evolution, and unbiased estimators, simple expressions are obtained for the quantum Cram\'er-Rao bound. This is not so, however, for open systems, that is, systems in the presence of an environment. Exact solutions can be found for one or two qubits \cite{fugiwara1,fugiwara2}, but for higher dimensions it is not possible, in general, to find analytical solutions. Lower bounds for the variance can be found through purification of the non--unitary dynamics, by adding an ad hoc environment, such that the dynamics of the enlarged system is unitary and the reduced dynamics, obtained by tracing out the added environment, coincides with the original dynamics of the system \cite{bruno,latune,taddei,variational}. Lower bounds for the precision have also been obtained via tools based on the geometry of quantum channels and semi-definite programming \cite{rafa4}. Also, exact solutions for the Cram\'er-Rao bound can be found for Gaussian systems \cite{paris2,banchi,nichols,dominik}.

Parameter estimation is closely related to quantum channel identification, that is, the distinguishability of quantum channels upon a change of one or more of the parameters defining the channel \cite{helstrom}. 
In quantum information, a quantum channel is a completely positive trace-preserving map between spaces of operators, where a map $\Gamma$ acting on operators in a Hilbert space 
${\cal H}_1$ is completely positive if the map 
$\Gamma\otimes I$ is positive when acting 
on all possible extensions ${\cal H}_1\otimes{\cal H}_2$ of ${\cal H}_1$. 
It is known that entanglement of the probe with an ancilla, with the channel acting only on the probe, may improve parameter estimation and the discrimination of quantum channels  \cite{fugiwara1,fugiwara2,acin,ariano,sacchi,wang,rafa2,rafa3,huang,pirandola}. 
The quantum advantage of this strategy is not universal, but it was demonstrated in some specific examples. In particular, it does not hold for unitary channels. Error correction, through the addition of multiple ancillas, has also been shown to increase the precision of estimation \cite{dur,dorit,kessler}.

Here we consider the quantum sensing of photon loss, due to absorption and scattering by a material {\cite{paris2,monras,gong,adesso,gammaT,moreau,losero,fli}}. It has direct application to the estimation of the transmissivity of light by a sample, especially for weak losses,  and when low intensities are desirable, which may be the case for fragile materials. {Quantum metrology of absorption is also important in absorption imaging. Reference \cite{brida} demonstrated sub-shot noise quantum imaging using entangled photons produced by a down converter. Reference \cite{zhang} demonstrated that the use of an ancilla in quantum illumination increases the signal-to-noise ratio beyond the classical value, in an entanglement-breaking environment.}

The absorption constant $\alpha$, to be estimated, is defined so that, if $I_0$ and $I_1$ are the intensities of light before and after the absorbing sample, then $I_1 = (1 - \alpha)I_0$.

While we concentrate here on the estimation of the absorption from a single-mode probe, this strategy is of broader application. For instance, it can be extended to the important spectroscopic technique that determines differential absorption of two orthogonal polarized modes, which has recently been investigated \cite{jiaxuan,nair}. 

\begin{figure}[b]
\centering\includegraphics[width=\columnwidth]{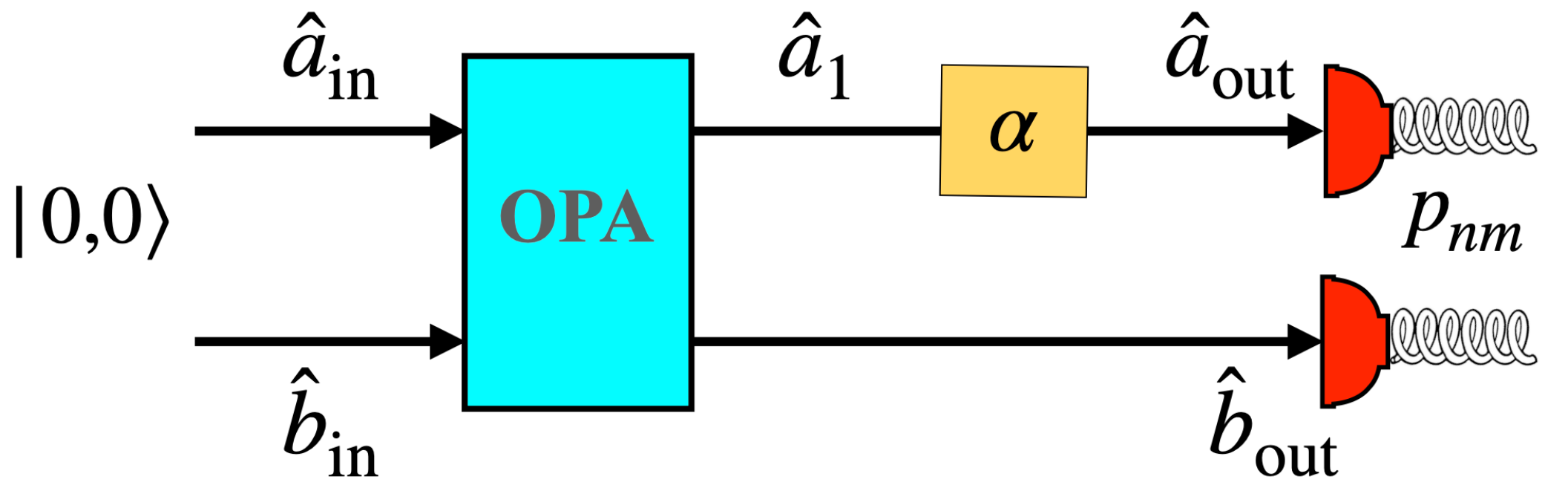}
\caption{Experimental setup for attaining a precision of estimation of an absorption coefficient $\alpha$ equivalent to the one obtained by a Fock state, using however a bimodal squeezed state produced by an optical parametric amplifier as input. The estimation of $\alpha$ is obtained from the joint photon counting of the two outgoing modes -- probe plus ancilla.}
\label{Fig1}
\end{figure}

\section{{Reaching the ultimate precision limit}}

The ultimate precision limit for the estimation of absorption of a sample in the setup considered here is obtained through the quantum Fisher information for the system probe + ancilla, which corresponds to the signal and idler beams of a bimodal squeezed state, produced by an optical parametric amplifier (OPA). Since the input state is Gaussian, the techniques used in \cite{paris2,banchi,nichols,dominik} can be applied to this case. This is done in Appendix A. Here, we adopt, however, a different procedure, which clarifies the physical meaning of our results, and leads to the QFI derived in the Appendix.

The system to be considered is pictured in Fig.~\ref{Fig1}. The OPA implements a bimodal squeezing transformation $\hat{S}(\xi)$ on a vacuum field \cite{agarwal2},
\begin{equation}\label{squeezing}
\hat S(\xi)= \exp(\xi\hat a^\dagger\hat b^\dagger-\xi^*\hat a\hat b)\,
\end{equation}
leading to the squeezed state
\begin{equation}\label{squeezed}
|\xi\rangle=\hat S(\xi)|0,0\rangle=\sum_{n=0}^\infty c_n|n,n\rangle\,,
\end{equation}
where $c_n= e^{in\varphi}(\tanh r)^n/\cosh r$. The average number of photons in either of the two modes is $\langle n\rangle=\sinh^2r$. 

The probability $p_{nm}$ of finding $n$ photons in the signal mode and $m$ photons in the idler mode, before the sample, is thus $p_{nm}=\delta_{nm}|c_n|^2$, while the probability of finding $m$ photons in the ancilla is $|c_m|^2$. After the sample is tested, the joint probability is $p_{nm}=|c_m|^2p_n^{(m)}$, where $p_n^{(m)}$ is the probability of counting $n$ photons in the output probe beam for the input of $m$ photons. The corresponding Fisher information for the estimation of the absorption $\alpha$ is then, according to Eq.~(\ref{fisher}), since only $p_n^{(m)}$ depends on $\alpha$,
\begin{eqnarray}\label{eqfisher}
F(\alpha)&=&\sum_{m=0}^\infty\sum_{n=0}^m \frac{1}{|c_m|^2p_n^{(m)}} \left[\frac{\partial |c_m|^2p_n^{(m)}}{\partial\alpha}\right]^2\nonumber\\
&=&\sum_{m=0}^\infty |c_m|^2\sum_{n=0}^m\frac{1}{p_n^{(m)}} \left[\frac{\partial p_n^{(m)}}{\partial\alpha}\right]^2=\sum_{m=0}^\infty|c_m|^2F^{(m)}(\alpha)\,,
\end{eqnarray}
where $F^{(m)}(\alpha)$ is the Fisher information for a $m$-photons Fock state probing the absorbing sample, obtained through photon counting. One knows, however, that photon counting actually leads to the quantum Fisher information for a Fock state input \cite{adesso,gammaT}, so $F^{(m)}(\alpha)={\cal F}_Q^{(m)}(\alpha)=m/[\alpha(1-\alpha)]$. Therefore, $F(\alpha)$ is a weighted average of QFIs corresponding to Fock states:
\begin{equation}\label{fisheralpha}
F(\alpha)={\sum_{m=0}^\infty|c_m|^2m\over\alpha(1-\alpha)}={\bar{n}\over\alpha(1-\alpha)}\,,
\end{equation}
since the sum in the first term on the right-hand side of the above equation is the average number of photons $\bar{n}$ in either the probe or the ancilla, before probing the sample. {For the incoming two-mode squeezed state considered here, $\bar n=\sinh^2r$}. In Appendix A Eq.~(\ref{qfi2}), it is shown that the Fisher information above coincides with the QFI of the probe + ancilla output state. Therefore,
\begin{equation}\label{fisherQalpha}
{\cal F}_Q(\alpha)={{\bar{n}\over\alpha(1-\alpha)}}\,.
\end{equation}

The Fisher information (\ref{fisheralpha}) coincides with the upper bound, derived in \cite{paris2}, on the QFI for the estimation of absorption, for any single-mode quantum state with mean photon number $\bar{n}$. This was done by replacing the absorption medium by a beam splitter, with transmissivity equal to the absorption coefficient, thus turning the open system dynamics into a unitary one, involving the two modes of the beam splitter. The corresponding quantum Fisher information should be an upper bound on the corresponding quantity for the open system, since having access to the environment should result in better precision of estimation of the absorption parameter \cite{bruno}.  Consequently, no single mode quantum state with a mean photon number $\bar{n}$ can beat the precision reached with a bimodal squeezed vacuum state with the same mean photon number in the probe beam. This can be generalized to the system probe plus ancilla considered here. If one replaces the absorbing medium by a beam splitter, as done for the single-mode case, the system will have a unitary evolution, and therefore the ancilla would not play any role: the upper bound is the same as in the single-mode case! And it is reached by the the two-mode squeezed state considered here, when the probe plus ancilla output is detected through joint photon counting.

{Eq.~(\ref{fisherQalpha}) leads to two important conclusions:}  
\begin{itemize}
\item[(i)] Joint photon counting on probe and ancilla is an optimal measurement, leading to the QFI corresponding to the parameter $\alpha$; 
\item[(ii)] The QFI related to the bimodal squeezed input state coincides with a Fock state QFI for which the photon number is replaced by $\bar n$. 
\end{itemize}
The resulting bound for the precision $\Delta \alpha$ in the estimation is given by Eq.~(\ref{qfi}):
\begin{equation}\label{qfifinal}
\Delta\alpha =\sqrt{{\alpha(1-\alpha)\over\bar n}}\,,
\end{equation}
{setting ${\cal N}=1$ in Eq.~(\ref{qfi}).}

As shown in \cite{adesso,gammaT}, Fock states lead to the best precision in the estimation of the absorption, for a fixed photon number. This implies that, through the use of an ancillary system, and for a given average photon number of photons probing the sample, it is possible to achieve the best precision in the estimation of the parameter $\alpha$,  outperforming not only the Gaussian states. 

 \begin{figure}[htbp]
\centering\includegraphics[width=\columnwidth]{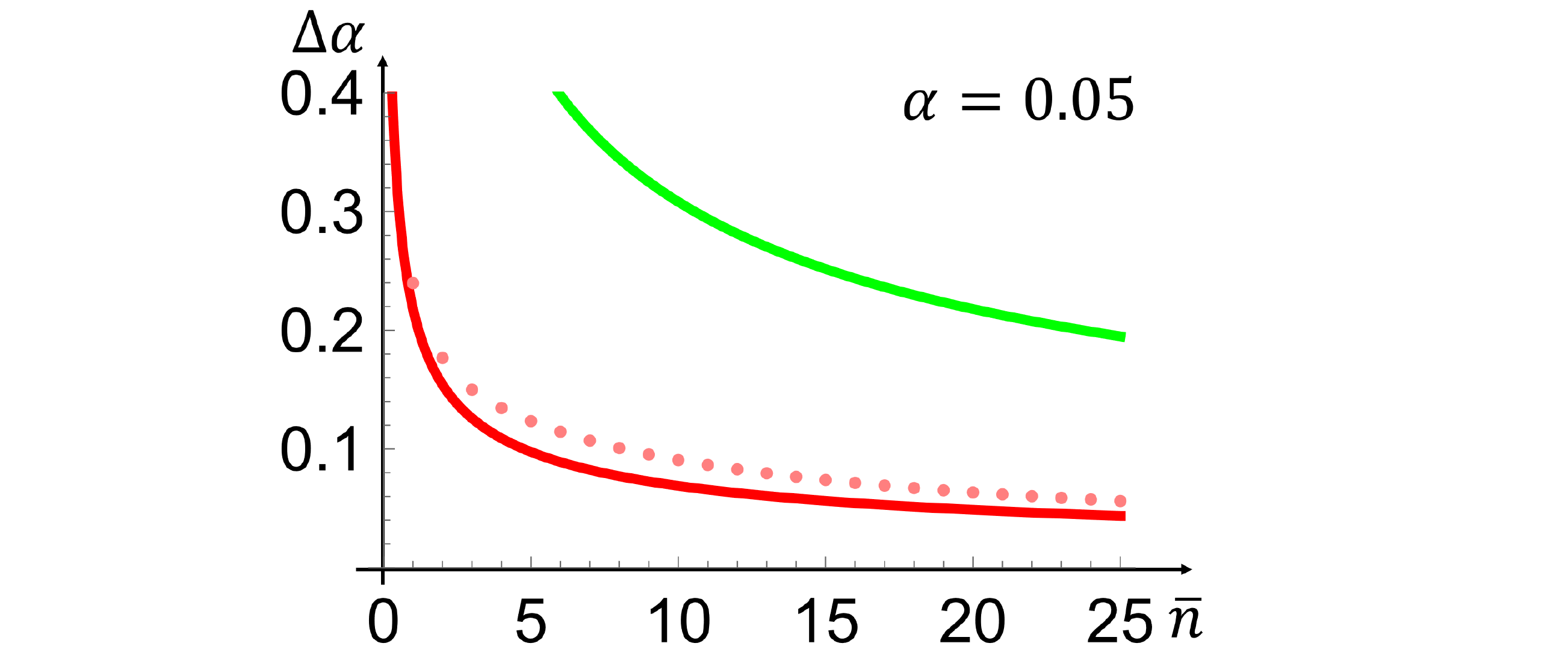}
\caption{Uncertainty in the estimation of the absorption constant, for  $\alpha=0.05$. Comparison between the bound for the uncertainty $\Delta\alpha$ obtained from the quantum Fisher information for probe plus ancilla corresponding to two modes of an incoming bimodal squeezed state (red line) and the bound from the best single-mode Gaussian state \cite{paris2} (pink dots), for the same average number of photons testing the absorption. The green line represents the standard limit, obtained for a single-mode coherent state testing the sample. For the probe plus ancilla setup with $\bar n=10$ the increase in precision from the standard limit is about 5 dB.} \label{Fig2}
\end{figure}

This interesting result stems from the perfect correlation of photon numbers of probe and ancilla, as shown in Eq.~(\ref{squeezed}).  Although the bimodal squeezed state describing the ancilla+probe system does not have a well-defined number of photons, photon counting on the ancilla tells us that a Fock state with the same number of photons probed the sample. The result of photon counting in the probe then allows one to get information about the parameter $\alpha$ as if the input state of the probe was a Fock state, leading to the best possible precision in the estimation of $\alpha$.  Several runs of joint photon counting lead to quantum Fisher information of a Fock state for which the photon number is replaced by $\bar n$. 

\subsection{Resilience of the joint photon counting procedure}

{The discussion above, {stemming from the derivation of the Fisher information in Eq.~(\ref{eqfisher}),} leads to important consequences: Any state, pure or mixed, with photon-number correlation between probe and ancilla could lead to Eq.~(\ref{qfifinal}). Not even entanglement is needed. Indeed, the following mixed product state of probe and ancilla,
\begin{equation}
\hat\rho=\sum_{n}p_n|n,n\rangle\langle n,n|
\end{equation}
has the same QFI as a two-mode vacuum squeezed state with the same average number of photons. This implies that the preparation of the input state via OPA is resilient to phase noise. Furthermore, it also allows one to understand and generalize numerical results published in \cite{monras}, where it was shown that the state $|\psi\rangle_d=(1/d)\sum_{k=1}^d|k,k\rangle$ with $3\le d\le 6$ has the same QFI as a two-mode vacuum squeezed state with the same average number of photons, and that there are states with less entanglement than  $|\psi\rangle_d$ with similar performance. }

\subsection{Quantum advantage of the probe+ancilla setup}
 
 Fig.~\ref{Fig2} compares the result for $\Delta\alpha$ obtained from the {ancilla-based} QFI Eq.~(\ref{fisherQalpha})  with the one corresponding to the best single-mode Gaussian state \cite{paris2}, which is a parameter-dependent squeezed and displaced vacuum state, for the same average number of photons probing the sample. The bound obtained from Eq.~(\ref{fisheralpha}) prevails, as expected from a QFI for a Fock-state expression with photon number equal to $\bar{n}=\sinh^2r$.  {This result is in conformity with \cite{monras}, which pointed out the non-optimal nature of single-mode Gaussian states.} One should note that both the input state and the detection procedure in the probe plus ancilla setup do not depend on the (unknown) parameter to be estimated, which is not the case of the procedure in \cite{paris2}.  
 
 In Fig.~\ref{Fig2}, these results are compared with the standard limit, which corresponds to probing the sample with an incoming coherent state, with the same average number of photons as in the previous setups, and measuring the intensity of the field after its interaction with the sample. It can be derived from the corresponding single-mode quantum Fisher information \cite{paris2}: 
 \begin{equation}\label{eqclassical}
 \Delta\alpha=\sqrt{{1-\alpha\over \bar n_1}}\,,
 \end{equation}
{where $\bar n_1$ is the incoming average number of photons probing the sample.} One should note that, for $\alpha\rightarrow1$, that is, for strong absorption, one has $\Delta\alpha\rightarrow0$. This is also the limit of vanishing {outgoing} intensity. Fig.~\ref{Fig2} shows that, for $\alpha=0.05$ and $\bar{n}=10$, the increase in precision from the standard limit is about 5 dB.  {In the limit of strong absorption, $\alpha\rightarrow1$, both quantum Fisher information, for the single mode and the probe plus ancilla setups, converge to} the standard limit, expressing the environment-induced emergence of classicality \cite{zurek,luiz}.  {This can be verified by comparing Eq.~(\ref{qfifinal}) and Eq.~(\ref{eqclassical}) when $\alpha\rightarrow1$.}

\section{Time-reversal strategy}

One should note, however, that joint photon counting is challenging, with present technologies. We show now that, for weak absorption, there is an interesting and useful alternative, which does not rely on joint photon counting, and involves a time-reversal detection scheme, illustrated in Fig.~\ref{Fig3}. It is based on a SU(1,1) interferometer  \cite{yurke,plick,lett,junxin,stuart,manceau,shengshuailiu,yuhongliu,jianqin,adhikari,yuhongliu2,liangcui,zyou,nanhuo,kalash},  consisting of two OPAs, with the probed sample between them. The first one generates a two-mode squeezed state from a vacuum input, with the signal mode probing the sample, and the idler beam playing the role of an ancilla. The second OPA reverses the transformation implemented by the first one, so that, in the absence of photon losses, there is no outgoing field. 
{The time-reversed operation can be carried out in many different ways \cite{shengshuailiu,yuhongliu,jianqin}. The simplest is to have a $\pi$ phase difference between the beams pumping the first and the second OPA.} We assume that proper calibration compensates for the difference in optical paths of both arms due to the presence of the sample. Detection of the total number of outgoing photons leads to the estimation of the absorption, defined{, as before,} by the constant $\alpha$, so that, if $I_1$ and $I_2$ are the intensities of light in the upper arm of the interferometer, before and after the absorbing medium, then $I_2=(1-\alpha)I_1$. {Fig.~\ref{Fig3}} displays the annihilation operators corresponding to the electromagnetic fields in several regions of the device. The relations between them are obtained from the squeezing transformations and the absorption. Thus, from the first OPA, one has \cite{agarwal2}
\begin{eqnarray}\label{opa1}
&&\hat{a}_1=\hat{a}_{\rm in}\cosh \!r+\hat{b}_{\rm in}^\dagger e^{i\phi}\sinh \!r\,,\nonumber\\
&&\hat{b}_1=\hat{b}_{\rm in}\cosh\! r+\hat{a}_{\rm in}^\dagger e^{i\phi}\sinh\! r\,,
\end{eqnarray}
where $[\hat a_{\rm in},\hat a_{\rm in}^\dagger]=1$, $[\hat b_{\rm in},\hat b_{\rm in}^\dagger]=1$ and the squeezing transformation is
\begin{equation}\label{s}
\hat{S}(\xi)=\exp(\xi\hat{a}^{\dagger}\hat{b}^\dagger-\xi^*\hat a\hat b)\,,
\end{equation}
where $\xi=re^{i\phi}$ is the squeezing parameter, with $\hat a_1=\hat{S}^{-1}\hat{a}_{\rm in}\hat S$, $\hat b_1=\hat{S}^{-1}\hat{b}_{\rm in}\hat S$.

The second OPA applies the time-reversed transformation $(\xi\rightarrow-\xi)$, resulting in the output operators {(see Fig.~\ref{Fig3}):}
\begin{eqnarray}\label{opa2}
&&\hat{a}_{\rm out}=\hat{a}_{2}\cosh \!r-\hat{b}_{1}^\dagger e^{i\phi}\sinh \!r\,,\nonumber\\
&&\hat{b}_{\rm out}=\hat{b}_1\cosh\! r-\hat{a}_{2}^\dagger e^{i\phi}\sinh\! r\,.
\end{eqnarray}

\begin{figure}[t]
\centering\includegraphics[width=8cm]{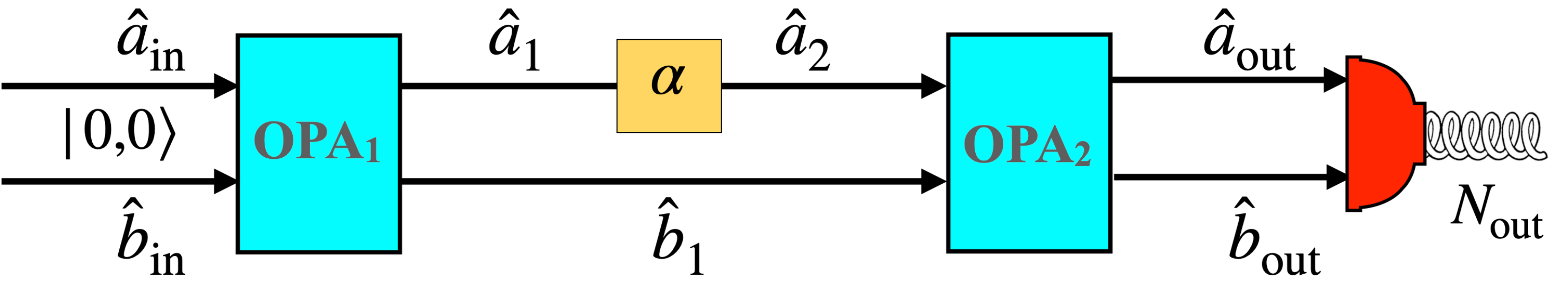}
\caption{Experimental SU(1,1) setup for squeezed vacuum time-reversal metrology. The absorption medium is placed between two optical parametric amplifiers (OPAs), on the upper arm of the interferometer. The first one, with vacuum input, produces a two-mode squeezed state, the signal beam probing the medium and the idler playing the role of an ancilla. The second OPA reverses the squeezing transformation, so that in the absence of the absorption medium, there is no outgoing field. Detection of the total number of outgoing photons leads to the estimation of the photon-loss coefficient $\alpha$.}
\label{Fig3}
\end{figure}

The photon loss, due to absorption and scattering, can be described by
\begin{equation}\label{absorber}
\hat{a}_2=\hat{a}_1\sqrt{1-\alpha}+\hat c\sqrt{\alpha}\,,
\end{equation}
where $\hat c$ stands for the annihilation operator corresponding to the vacuum noise mode. The presence of $\hat c$ preserves the commutation relation of the field operators: $[\hat{a}_2,\hat{a}^\dagger_2]=[\hat{a}_1,\hat{a}^\dagger_1]=1$.

From Eq.~(\ref{opa1}), Eq.~(\ref{opa2}), and Eq.~(\ref{absorber}), it follows that
\begin{eqnarray}\label{inout}
\hat{a}_{\rm out}&=&\hat{a}_{\rm in}\left(\cosh^2r\sqrt{1-\alpha}-\sinh^2r\right)\nonumber\\
&-&\hat{b}_{\rm in}^\dagger e^{i\phi}\!\sinh\!r\cosh\!r\!\left(1-\sqrt{1-\alpha}\right)+\hat{c}\cosh\!r\sqrt{\alpha}\,,\nonumber\\
\hat{b}_{\rm out}&=&\hat{b}_{\rm in}\left(\cosh^2r-\sinh^2r\sqrt{1-\alpha}\right)\nonumber\\
&+&\hat{a}_{\rm in}^\dagger e^{i\phi}\!\sinh\!r\cosh\!r\!\left(1-\sqrt{1-\alpha}\right)\nonumber\\
&-&\hat{c}^\dagger e^{i\phi}\sinh\!r\sqrt{\alpha}\,.
\end{eqnarray}
In the absence of the sample, it is easy to check that $\hat{a}_{\rm out}=\hat{a}_{\rm in}$, $\hat{b}_{\rm out}=\hat{b}_{\rm in}$.

From Eq.~(\ref{inout}), one gets the average total number of output photons:
\begin{eqnarray}\label{total}
\bar{N}_{\rm out}&=&\langle\hat a_{\rm out}^\dagger\hat a_{\rm out}+\hat b_{\rm out}^\dagger\hat b_{\rm out}\rangle \nonumber\\
&=&2\sinh^2\!r\cosh^2\!r(1-\sqrt{1-\alpha})^2+\alpha\sinh^2r\,.
\end{eqnarray}
The variance $\Delta^2N_{\rm out}$ is displayed in Appendix B Eq.~(\ref{noisesimp}).

We calculate $\Delta\alpha$ through the sensitivity, which can be related in this case to photon number fluctuations:
\begin{equation}\label{sensitivity}
\Delta\alpha={\Delta N_{\rm out}\over\left|d\bar{N}_{\rm out}/d\alpha\right|}\,,
\end{equation}
where
\begin{equation}\label{varianceN}
 \Delta^2 N_{\rm out}=\langle (N_{\rm out}-\bar{N}_{\rm out}\rangle)^2\rangle
 \end{equation}
 is the variance of the total number of photons. 
 
 \begin{figure}[b]
\centering\includegraphics[width=8cm]{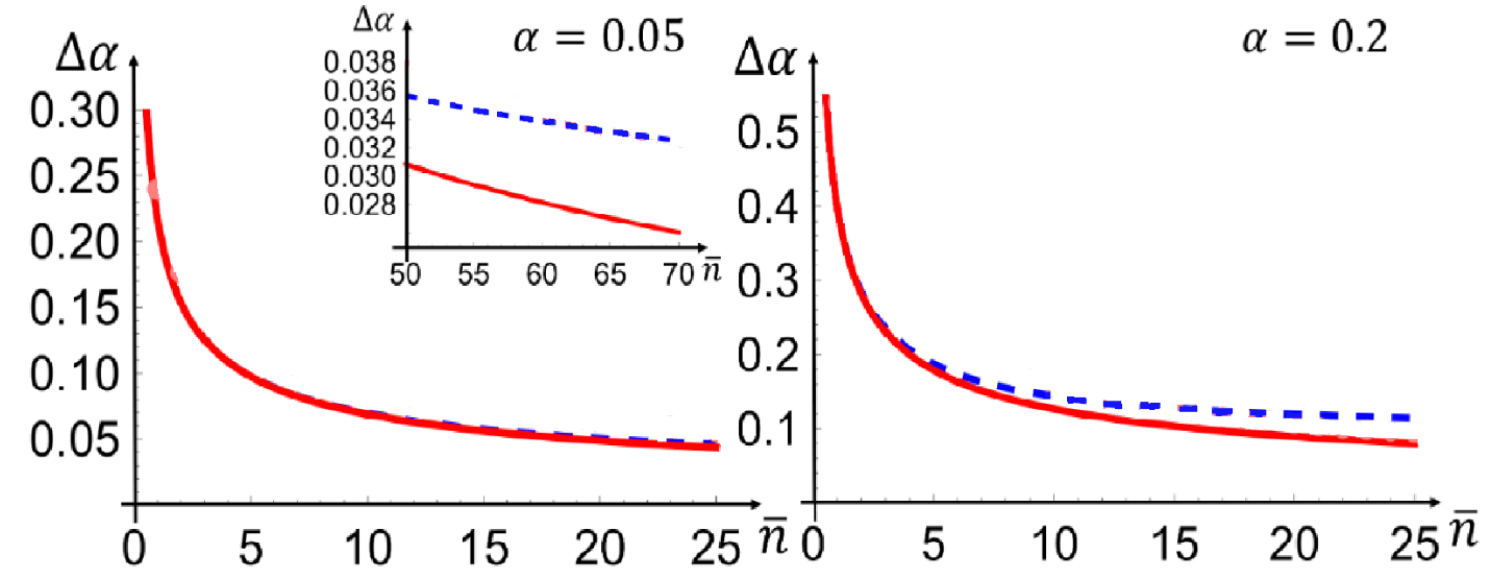}
\caption{Uncertainty bounds for probe plus ancilla QFI and SU(1,1) result. The uncertainty bounds from the quantum Fisher information corresponding to the probe and ancilla associated with the two modes of a bimodal squeezed state (red curve), given by Eq.~(\ref{fisheralpha}), and the one resulting from a sensitive calculation for the total number of outgoing photons in the SU(1,1) setup illustrated in Fig.~\ref{Fig3} (dashed blue line). For weak absorption ($\alpha\ll 1$), results are practically indistinguishable for the range of $\bar{n}$ considered here.}
\label{Fig4}
\end{figure}

From these expressions, the sensitivity can be calculated. Details are given in Appendix B Eqs.~(\ref{sensitivity2})\,-\,(\ref{B}). The corresponding uncertainty is plotted in Fig.~\ref{Fig4}, and compared with the one obtained from the QFI in Eq.~(\ref{fisherQalpha}). For weak absorption, the result obtained from the time-reversal procedure is practically indistinguishable from the probe plus ancilla quantum Fisher information bound.

Since only the measurements of the total output photon number and its variance are needed here, they can be obtained through measurement of the intensity of photocurrents produced by the output fields and their cross-correlations, which does not require reconstructing the photon-number distribution \cite{lmandel}.

In the next section, we demonstrate another advantage
of this method: the resilience to moderate photon losses
of the incoming probe plus the ancilla beam.

\section{Resilience of time-reversal to extra photon losses}

\begin{figure}[tb]
\centering
\includegraphics[width=\columnwidth]{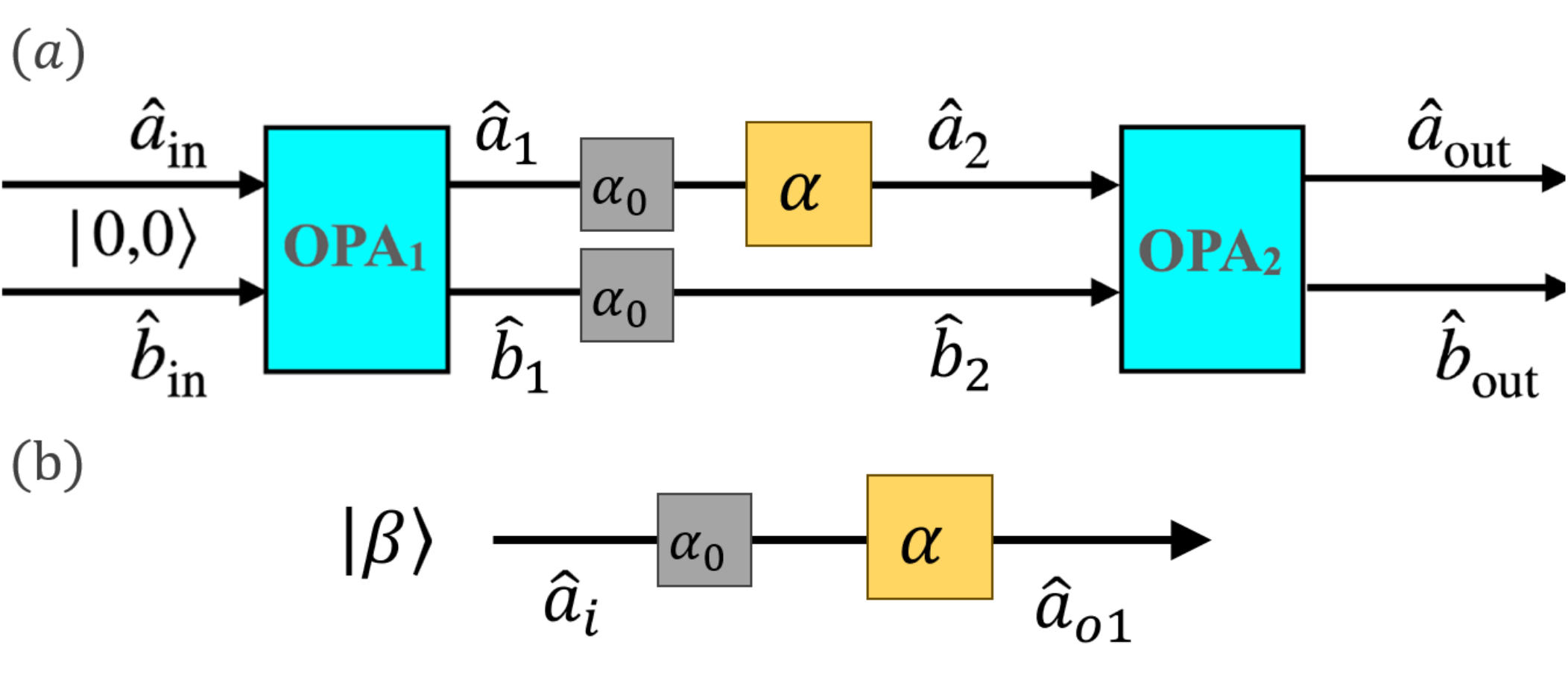}
\caption{(a). The impurity of the two-mode squeezed state generated by general optical parametric amplifiers is modeled by adding additional loss to the two-mode squeezed vacuum before passing through the sample $\alpha$. (b). For comparison, the corresponding sensing protocol with a coherent beam is shown.}
\label{Fig5}
\end{figure}

\begin{figure}[tb]
\centering
\includegraphics[width=0.8\columnwidth]{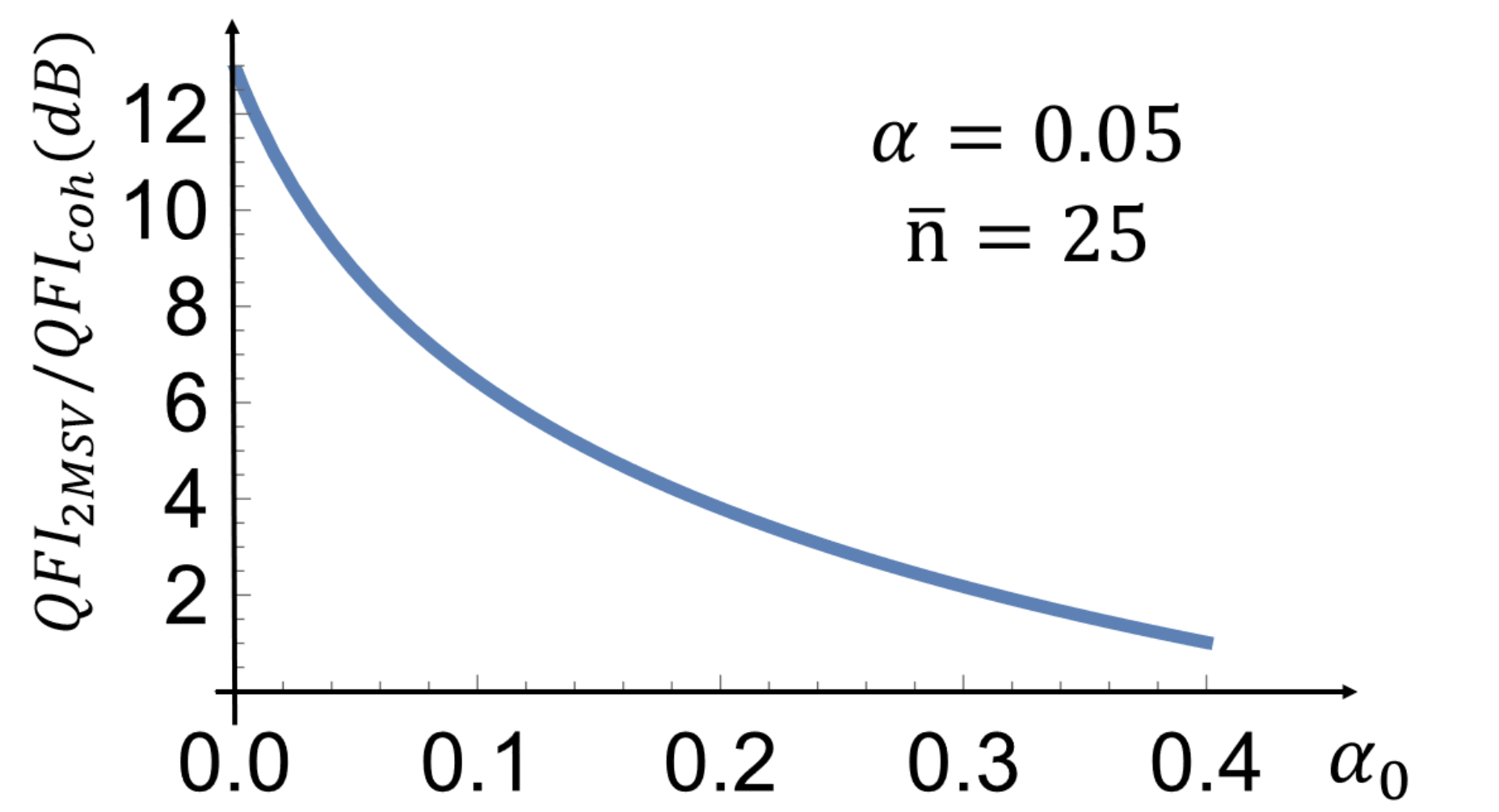}
\caption{The ratio between the quantum Fisher Informations corresponding to a  two-mode squeezed vacuum (2MSV) and to a coherent state, $QFI_{2MSV}/QFI_{coh}$, in dB, for $\alpha=0.05$ and $\bar{n}=25$, as a function of the impurity loss $\alpha_0$.}
\label{Fig6}
\end{figure}

General optical parametric amplifiers (OPAs) can produce impure entanglement due to various factors, leading to a reduction in the quality of the generated entanglement. We have already shown that the joint photon counting procedure is not affected if the input state becomes a mixture of photon-correlated probe plus ancilla states. We consider now the effect of extra photon losses.

It is worth noting that while losses can introduce impurities in the entanglement produced by general OPAs, it is still possible to have quantum gain for the joint photon counting and the time-reversal setup. In Fig.~\ref{Fig5}, additional loss is introduced to both modes of the two-mode squeezed vacuum before it passes through the sample. For simplicity, we assume the degree of loss $\alpha_0$ to be the same for both modes. The calculations are similar to those in Appendix B.

As shown in Fig.~\ref{Fig6}, where the QFI for estimation of  $\alpha$ corresponding to the two-mode squeezed state is compared to the QFI for a single-mode coherent state input with the same additional loss $\alpha_0$, quantum advantage persists even after undergoing significant loss (advantage of 3~dB with $\alpha_0\sim5\alpha$). This actually refers to the joint photon counting procedure, since the second OPA does not change the QFI.

The resilience of the time-reversal procedure to noise is illustrated in Fig.~\ref{Fig7}. Even for extra noise equal to the sample absorption constant, there is still significant advantage, as compared to the estimation corresponding to a coherent state input, subject to the same extra noise. This is an important and useful property of the time-reversal strategy proposed here.

\begin{figure}[tb]
\centering
\includegraphics[width=\columnwidth]{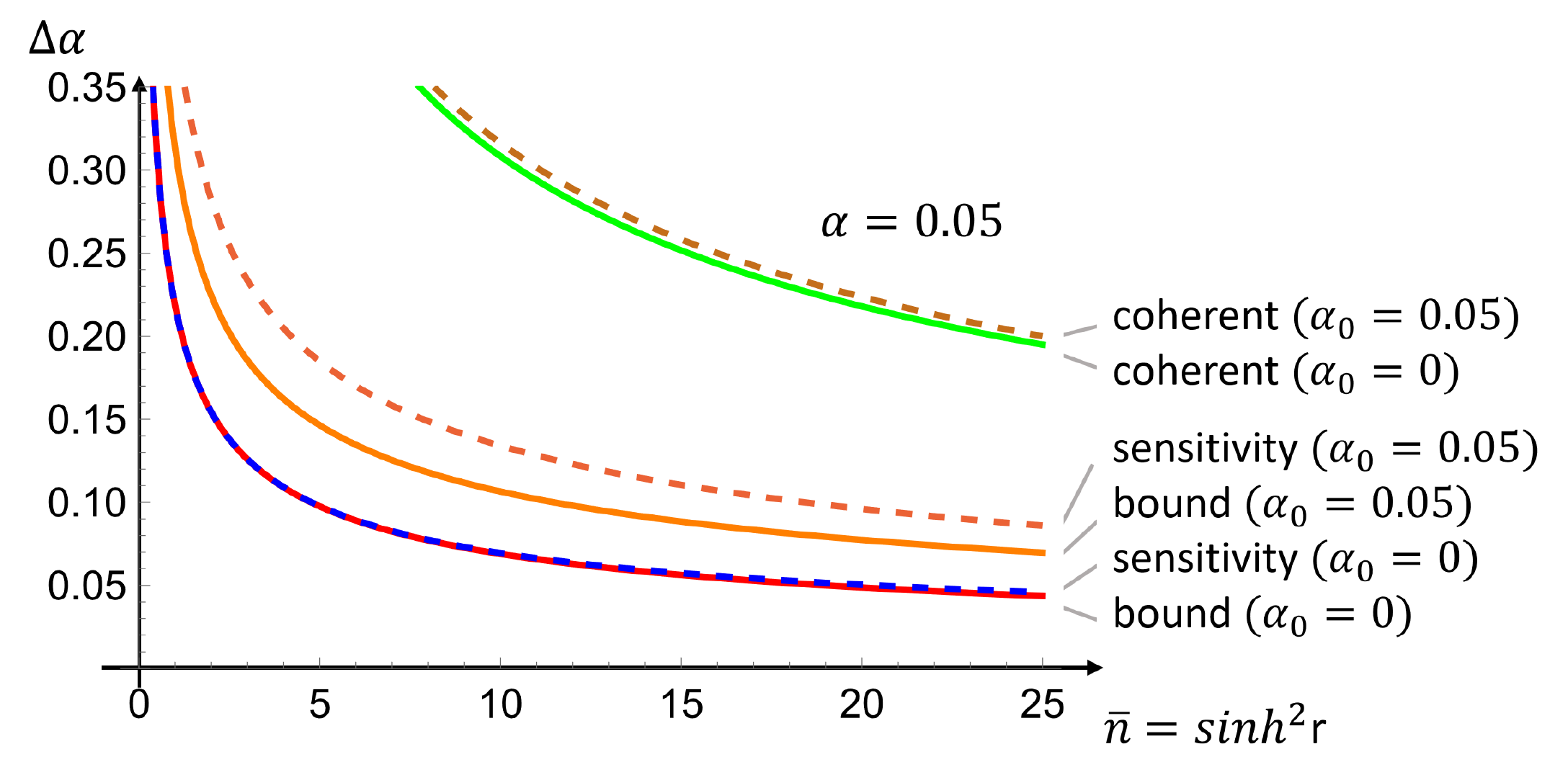}
\caption{Resilience of the time-reversal procedure:  Precision $\Delta\alpha$ in the estimation of the sample absorption constant as a function of the extra loss $\alpha_0$ in both probe and conjugate beams, as in Fig.~\ref{Fig5}(a). For comparison, the standard quantum limit, corresponding to a single-mode coherent state input, as in Fig.~\ref{Fig5}(b),  is also shown. The scale on x axis is the number of photons produced by the OPA in each beam.}
\label{Fig7}
\end{figure}

\section{Discussion}

Optimal quantum sensing of open-system dynamics may require strategies that differ markedly from those applied to lossless systems. Entanglement of the probe with an ancilla may enhance the precision of estimation, even though the ancilla does not interact with the parameter-dependent system, a property that is absent for unitary dynamics.  {Here we have considered the estimation of photon loss for a light beam propagating in a sample or, more generally, of the loss coefficient in a bosonic channel, due to absorption or scattering by the sample.}

The probe and ancilla are the modes of a bimodal squeezed state, produced by an optical parametric amplifier. We calculated the corresponding quantum Fisher information for estimation of the absorption coefficient and showed that the respective uncertainty bound coincides with the one for a Fock state, being saturated by joint photon counting for the outgoing probe and ancilla, a detection procedure that does not depend on the value of the parameter. The ancilla strategy benefits therefore from the extreme precision associated with Fock states of the probe, surpassing the sensing obtained with the best parameter-dependent single-mode Gaussian state \cite{paris2}, while overcoming the challenge of producing high photon-number eigenstates.

{Joint photon counting is however challenging, with present technologies \cite{nair1}. We have therefore described an approach that does not require this procedure.} It is based on the conjunction of the ancilla strategy with a time-reversal strategy, implemented with a SU(1,1) interferometer consisting of two optical parametric amplifiers, so that the first one generates a two-mode squeezed state, corresponding to the probe and the ancilla, and the second one undoes the squeezing produced by the first one, after the probe has interacted with the sample. The addition of the second optical parametric amplifier does not change the quantum Fisher information, which is invariant under unitary transformations. An estimation based on a sensitivity relation for the total number of outgoing photons and its variance leads to an uncertainty in the absorption coefficient with a precision that is, for weak absorption, practically indistinguishable from the bound obtained from the quantum Fisher information for the probe plus ancilla system. This detection setup is also independent of the (unknown) parameter to be estimated, implying that phase stabilization and mode-matching for the two optical parametric amplifiers can be done once and for all: the device is then ready to be used, independently of the value of the parameter.  {Interestingly, quantum gain is still achieved for this protocol under moderate photon losses of the probe plus ancilla input beam, even for losses compared to the estimated absorption parameter.}

{The quantum advantage in the precision of estimation obtained with the ancilla and time-reversal strategy, demonstrated here, relies on available technology and opens the way to the increase in precision of a diversity of metrological tasks involving open systems.}

\section*{Appendix A. quantum Fisher information of absorption constant for the time-reversed ancilla-based metrology}
\renewcommand{\theequation}{A\arabic{equation}}
\setcounter{equation}{0}

We note that the two-mode squeezed state has a Gaussian Wigner function. It can be shown that if one of the modes goes through an absorber, the output is still a Gaussian Wigner function, though it corresponds to a mixed state of the output field \cite{agarwal1987wigner}.
The general method for calculating the quantum Fisher information (QFI) for a Gaussian system has been developed by \cite{pinel2012ultimate, pinel2013quantum, gao2014bounds, friis2015heisenberg, vsafranek2015quantum, banchi2015quantum, marian2016quantum, nichols2018multiparameter, vsafranek2018estimation}.
We apply specifically the method in \cite{vsafranek2018estimation} to obtain the quantum Fisher information of the system described by Fig.~1. 
For a two-mode bosonic system, we may define a vector of annihilation and creation operators given by
\begin{align}\label{a}
\hat{A}=\left(
a_1,
a_2,
a_1^{\dagger},
a_2^{\dagger}
\right)^T\,.
\end{align}
Gaussian states can be fully characterized by their first moments (the displacement vector) $ \newcommand{\bd}{d} \newcommand{\bA}{{\hat{A}}} \newcommand{\bi}{\boldsymbol} \bd_m=\mathrm{tr}\big[\rho\bA_m\big]$ and the second moments (the covariance matrix) $ \newcommand{\e}{{\rm e}} \newcommand{\de}{{\rm d}\epsilon} \newcommand{\D}{\delta} \newcommand{\dbA}{{\Delta\hat{A}}} \newcommand{\re}{{\rm Re}} \newcommand{\bi}{\boldsymbol} \renewcommand{\dag}{\dagger}\sigma_{mn}=\mathrm{tr}\big[\hat{\rho}\,\{\dbA_m,\dbA_n^{\dag}\}\big]$, where $ \newcommand{\e}{{\rm e}} \newcommand{\de}{{\rm d}\epsilon} \newcommand{\D}{\delta} \newcommand{\bd}{d} \newcommand{\bA}{{\hat{A}}} \newcommand{\dbA}{{\Delta\hat{A}}} \dbA:=\bA-\bd$. The subscripts $m$ and $n$ stand for the components of the vector defined in Eq.~(\ref{a}). 
\noindent The QFI is given by
\begin{align}\label{qfi}
{\cal F}_Q(\alpha)=lim_{v\rightarrow1}\frac{1}{2}vec(\frac{\partial\sigma}{\partial\alpha})^{\dagger}M^{-1}vec(\frac{\partial\sigma}{\partial\alpha})\,,
\end{align}
where the matrix $M$ can be expressed by
\begin{align}\label{m}
M=v^{2}\bar{\sigma}\otimes\sigma-K\otimes K\,,
\end{align}
for the system considered here, with a zero displacement vector. Other notations in Eq.~(\ref{qfi}) include the symplectic form $K=diag(1,1,-1,-1)$, and the operator $vec(\Lambda)$. 
Applying $vec(\Lambda)$ on a matrix $\Lambda=\left(\Lambda_{1},\Lambda_{2}\right)$ will transform it to a vector $vec(\Lambda)=\left(\Lambda_{1}^{T},\Lambda_{2}^{T}\right)^{T}$.
Replacing the operators $a_1$ and $b_1$ in Eq.~(\ref{a}) by $\hat{a}_{\rm out}$ and $\hat{b}_{\rm out}$, we obtain
{\small
\begin{align}
\label{sigma}
\sigma
=\left(\begin{array}{cccc}
\left\langle \hat{a}_{o}\hat{a}_{o}^{\dagger}+\hat{a}_{o}^{\dagger}\hat{a}_{o}\right\rangle  & 2\left\langle \hat{a}_{o}\hat{b}_{o}^{\dagger}\right\rangle  & 2\left\langle \hat{a}_{o}\hat{a}_{o}\right\rangle  & 2\left\langle \hat{a}_{o}\hat{b}_{o}\right\rangle \\
2\left\langle \hat{a}_{o}^{\dagger}\hat{b}_{o}\right\rangle  & \left\langle \hat{b}_{o}\hat{b}_{o}^{\dagger}+\hat{b}_{o}^{\dagger}b_{o}\right\rangle  & 2\left\langle \hat{a}_{o}\hat{b}_{o}\right\rangle  & 2\left\langle \hat{b}_{o}\hat{b}_{o}\right\rangle \\
2\left\langle \hat{a}_{o}^{\dagger}\hat{a}_{o}^{\dagger}\right\rangle  & 2\left\langle \hat{a}_{o}^{\dagger}\hat{b}_{o}^{\dagger}\right\rangle  & \left\langle \hat{a}_{o}^{\dagger}\hat{a}_{o}+\hat{a}_{o}\hat{a}_{o}^{\dagger}\right\rangle  & 2\left\langle \hat{a}_{o}^{\dagger}\hat{b}_{o}\right\rangle \\
2\left\langle \hat{a}_{o}^{\dagger}\hat{b}_{o}^{\dagger}\right\rangle  & 2\left\langle \hat{b}_{o}^{\dagger}\hat{b}_{o}^{\dagger}\right\rangle  & 2\left\langle \hat{a}_{o}\hat{b}_{o}^{\dagger}\right\rangle  & \left\langle \hat{b}_{o}^{\dagger}b_{o}+\hat{b}_{o}\hat{b}_{o}^{\dagger}\right\rangle 
\end{array}\right)\,,
\end{align}}
where for brevity, we note $\hat{a}_{\rm out}$ and $\hat{b}_{\rm out}$ by $\hat{a}_{\rm o}$ and $\hat{b}_{\rm o}$. These are given by
\begin{align}
\hat{a}_o&=\hat{a}_1\sqrt{1-\alpha}+\hat c\sqrt{\alpha}\,,\nonumber \\
\hat{b}_o&=\hat{b}_1,
\end{align}
where $\hat{a}_1$ and $\hat{b}_1$ are given by Eq.~(11). In the system considered here, many of the off-diagonal terms are zero
\begin{align}\label{zeros}
\left\langle \hat{a}_{o}\hat{a}_{o}\right\rangle &=\left\langle \hat{a}_{o}^{\dagger}\hat{a}_{o}^{\dagger}\right\rangle =\left\langle \hat{b}_{o}\hat{b}_{o}\right\rangle =\left\langle \hat{b}_{o}^{\dagger}\hat{b}_{o}^{\dagger}\right\rangle \nonumber \\
&=\left\langle \hat{a}_{o}\hat{b}_{o}^{\dagger}\right\rangle =\left\langle \hat{a}_{o}^{\dagger}\hat{b}_{o}\right\rangle =0\,.
\end{align}
\noindent Thus, we obtain
{\small
\begin{align}
\label{sigma2}
\sigma
=\left(\begin{array}{cccc}
2\left\langle \hat{a}_{o}^{\dagger}\hat{a}_{o}\right\rangle +1 & 0 & 0 & 2\left\langle \hat{a}_{o}\hat{b}_{o}\right\rangle \\
0 & 2\left\langle \hat{b}_{o}^{\dagger}b_{o}\right\rangle +1 & 2\left\langle \hat{a}_{o}\hat{b}_{o}\right\rangle  & 0\\
0 & 2\left\langle \hat{a}_{o}^{\dagger}\hat{b}_{o}^{\dagger}\right\rangle  & 2\left\langle \hat{a}_{o}^{\dagger}\hat{a}_{o}\right\rangle +1 & 0\\
2\left\langle \hat{a}_{o}^{\dagger}\hat{b}_{o}^{\dagger}\right\rangle  & 0 & 0 & 2\left\langle \hat{b}_{o}^{\dagger}b_{o}\right\rangle +1
\end{array}\right)\,,
\end{align}}
\noindent where
\begin{align}\label{c123}
\left\langle \hat{a}_{o}^{\dagger}\hat{a}_{o}\right\rangle &=(1-\alpha)\sinh^{2}r\,,\nonumber \\
\left\langle \hat{b}_{o}^{\dagger}b_{o}\right\rangle &=\sinh^{2}r\,,\nonumber \\
\left\langle \hat{a}_{o}\hat{b}_{o}\right\rangle &=\left\langle \hat{b}_{o}\hat{a}_{o}\right\rangle =e^{i\phi}\sinh r\cosh r\sqrt{1-\alpha}\,,\nonumber \\
\left\langle \hat{a}_{o}^{\dagger}\hat{b}_{o}^{\dagger}\right\rangle &=\left\langle\hat{b}_{o}^{\dagger}\hat{a}_{o}^{\dagger}\right\rangle =e^{-i\phi}\sinh r\cosh r\sqrt{1-\alpha}\,.\nonumber \\
\end{align}
\noindent 
From Eq.~(\ref{qfi}), Eq.~(\ref{zeros}), and Eq.~(\ref{sigma2}), we obtain the quantum Fisher information corresponding to the absorption constant $\alpha$, for the ancilla-based metrology,
\setlength{\belowdisplayskip}{7pt} \setlength{\belowdisplayshortskip}{7pt}
\setlength{\abovedisplayskip}{7pt} \setlength{\abovedisplayshortskip}{7pt}
\begin{align}\label{qfi2}
{\cal F}_Q(\alpha)=\frac{\sinh^{2}r}{\alpha(1-\alpha)}\,,
\end{align}
which coincides with Eq.$\,$(6) of the main text. {We note that this result can be derived from the quantum Fisher information obtained in \cite{monras}, where it was shown that input two-mode squeezed states outperform any other class of Gaussian states, for the estimation of the time-dependent parameter $\gamma=\Gamma t$,  where $\Gamma$ is the coupling of the channel to a thermal reservoir. The connection with Fock states, through the expression Eq.~(\ref{qfi2}), was not discussed there, as well as the resilience of these states against additional noise, which are direct consequences of our derivation method in the main text. Since the time-reversal procedure is a unitary transformation, which does not change the quantum Fisher information, Eq.~(\ref{qfi2}) also applies to the time-reversed ancilla system. This can be checked explicitly for the arrangement of Fig.~\ref{Fig3} for which $\hat{a}_{\rm out}$ and $\hat{b}_{\rm out}$ are given by Eq.~(\ref{inout}). We can derive

\begin{align}\label{c1232}
\left\langle \hat{a}_{o}^{\dagger}\hat{a}_{o}\right\rangle &=\sinh^{2}r\cosh^{2}r\left(1-\sqrt{1-\alpha}\right)^{2}\,,\nonumber \\
\left\langle \hat{b}_{o}^{\dagger}b_{o}\right\rangle &=\sinh^{2}r\cosh^{2}r\left(1-\sqrt{1-\alpha}\right)^{2}+\alpha\sinh^{2}r\,,\nonumber \\
\left\langle \hat{a}_{o}\hat{b}_{o}\right\rangle &=\left\langle \hat{b}_{o}\hat{a}_{o}\right\rangle =-e^{i\phi}\sinh r\cosh r\left(1-\sqrt{1-\alpha}\right) \nonumber \\
&\times\left(\cosh^{2}r-\sinh^{2}r\sqrt{1-\alpha}\right)\,,\nonumber \\
\left\langle \hat{a}_{o}^{\dagger}\hat{b}_{o}^{\dagger}\right\rangle &=-e^{-i\phi}\sinh r\cosh r\left(1-\sqrt{1-\alpha}\right)\nonumber \\
&\times\left(\cosh^{2}r-\sinh^{2}r\sqrt{1-\alpha}\right)\,.\nonumber \\
\end{align}
On substituting Eq.~(\ref{c1232}) in Eq.~(\ref{sigma2}) and using Eq.~(\ref{qfi}) and Eq.~(\ref{m}), we do obtain Eq.~(\ref{qfi2}).

\section*{Appendix B. sensitivity for the time-reversed scheme}
\renewcommand{\theequation}{B\arabic{equation}}
\setcounter{equation}{0}

Here we present the expressions needed for the evaluation of $\Delta\alpha$ in Eq.~(\ref{sensitivity}).
For simplicity, we express Eq.~(\ref{inout}) in the main text as
\begin{align}\label{c123}
\hat{a}_{\rm out}&=c_{11}\hat{a}_{\rm in}+c_{12}\hat{b}_{\rm in}^\dagger+c_{13}\hat{c}\,,\nonumber \\
\hat{b}_{\rm out}&=c_{21}\hat{a}_{\rm in}^{\dagger}+c_{22}\hat{b}_{\rm in}+c_{23}\hat{c}^{\dagger}\,,
\end{align}
where $c_{11}$, $c_{12}$, $c_{13}$, $c_{21}$, $c_{22}$, and $c_{23}$ are given by
\begin{align}\label{c123}
c_{11}&=\sqrt{1-\alpha}\cosh^{2}r-\sinh^{2}r\,,\nonumber \\
c_{12}&=-(1-\sqrt{1-\alpha})e^{i\phi}\cosh r\sinh r\,,\nonumber \\
c_{13}&=\cosh r\sqrt{\alpha}\,,\nonumber \\
c_{21}&=(1-\sqrt{1-\alpha})e^{i\phi}\cosh r\sinh r\,,\nonumber \\
c_{22}&=\cosh^{2}r-\sqrt{1-\alpha}\sinh^{2}r\,,\nonumber \\
c_{23}&=-e^{i\phi}\sinh r\sqrt{\alpha}\,.\nonumber \\
\end{align}
We obtain the average total number of output photons
\begin{align}\label{signal0}
\bar{N}_{\rm out}&=\langle\hat a_{\rm out}^\dagger\hat a_{\rm out}+\hat b_{\rm out}^\dagger\hat b_{\rm out}\rangle =|c_{12}|^{2}+|c_{21}|^{2}+|c_{23}|^{2}\,,
\end{align}
and its variance
\begin{align}\label{noise}
\Delta^2 N_{\rm out}=&\langle (N_{\rm out}-\bar{N}_{\rm out}\rangle)^2\rangle \nonumber \\
=&|c_{12}|^{2}(|c_{11}|^{2}+|c_{13}|^{2}+|c_{22}|^{2})+|c_{22}|^{2}|c_{23}|^{2}
\nonumber \\
+&2|c_{12}c_{22}(c_{11}c_{21}+c_{13}c_{23})|\,.
\end{align}
Defining $\eta:=(1-\sqrt{1-\alpha})$, we obtain
\begin{equation}\label{dsda}
\frac{d\bar{N}_{\rm out}}{d\alpha}=2\cosh^{2}r\sinh^{2}r\frac{\eta}{1-\eta}+\sinh^{2}r\,,
\end{equation}
and
\begin{align}\label{noisesimp}\Delta^2 N_{\rm out}&=\eta^{2}\cosh^{2}r(2\eta-1+3\eta^{2}\cosh^{2}r\sinh^{2}r)\nonumber \\
&+2\alpha\eta \cosh^{2}r(1+\eta \sinh^{2}r)\nonumber \\
&+(1+\eta \sinh^{2}r)^{2}(2\eta \cosh^{2}r-\alpha \sinh^{2}r)\,.
\end{align}
From these expressions, the sensitivity can be calculated. 

\begin{equation}\label{sensitivity2}
\Delta\alpha={\sqrt{A}\over B}\,,
\end{equation}
where
\begin{eqnarray}\label{A}
A&=&\eta^{2}\left(\bar{n}+1\right)\left[2\eta-1+3\eta^{2}(\bar{n}+1)\bar{n}\right]\nonumber\\
&+&2\alpha\eta\left(\bar{n}+1\right)\left(1+\eta\bar{n}\right)\nonumber\\
&+&\left(1+\eta\bar{n}\right)^{2}\left[2\eta(\bar{n}+1)-\alpha\bar{n}\right],
\end{eqnarray}
 and 
 \begin{equation}\label{B}
 B=\sqrt{\bar{n}}\left[1+\frac{2\eta}{\sqrt{1-\alpha}}\left(\bar{n}+1\right)\right],
 \end{equation}
in terms of the average number of photons interacting with the sample, $\bar{n}=\langle \hat a_1^\dagger \hat a_1\rangle=\sinh^2r$, and $\eta:=(1-\sqrt{1-\alpha})$.
{For small values of $\alpha$ such that $\alpha\ll1$ and $\alpha\bar{n}\lesssim 1$, Eq.~(\ref{sensitivity2}) can be expanded to
\begin{equation}\label{Aapx}
\Delta\alpha=\frac{\sqrt{\alpha}[1+\alpha n+\frac{1}{8}\alpha^{3}n^{3}+\frac{1}{2}\alpha(1+\frac{1}{2}\alpha n+\frac{1}{2}\alpha^{2}n^{2})]}{\sqrt{\bar{n}}\left[1+\alpha\bar{n}+\alpha\left(1+\frac{3}{4}\alpha\bar{n}\right)\right]},
\end{equation}
where the leading term is $\Delta\alpha\sim\sqrt{\alpha/\bar{n}}$, showing that the SU(1, 1) sensitivity estimate goes over the result in Eq.~(\ref{qfifinal}), for $\alpha\ll1$.}

\section{ACKNOWLEDGMENTS}
G.S.A and J.W. are grateful for the support of Air Force Office of Scientific Research (Award No. FA-9550-20-1-0366) and the Robert A. Welch Foundation (A-1943-20210327). R.L.M.F. and L.D. acknowledge the support of the Brazilian agencies CNPq, CAPES, and the Rio de Janeiro State Foundation for Research Support (FAPERJ). R. L. M. F. acknowledges the support of the John Templeton Foundation (Grant 62424). L.D. acknowledges the support by the National Science Foundation under Grants No. NSF PHY-1748958 and PHY-2309135.



\end{document}